# Direct Observation of Long Electron-Hole Diffusion Distance in $CH_3NH_3PbI_3$ Perovskite Thin Film

Yu Li[1, §], Weibo Yan[2, §], Yunlong Li[2], Shufeng Wang[1,*], Wei Wang[1], Zuqiang Bian[2,*], Lixin Xiao[1], and Qihuang Gong[1]

[1]Institute of Modern Optics & State Key Laboratory for Artificial Microstructure and Mesoscopic Physics, School of Physics, Peking University, Beijing 100871, China.

[2]State Key Laboratory of Rare Earth Materials Chemistry and Applications, College of Chemistry and Molecular Engineering, Peking University, Beijing, 100871, China

[§]These authors contributed equally to this work. *Correspondence and requests for materials should be addressed to S.W. (email: wangsf@pku.edu.cn) or to Z.B. (email: bianzq@pku.edu.cn)

**ABSTRACT**

In high performance perovskite based solar cells, $CH_3NH_3PbI_3$ is the key material. We carried out a study on charge diffusion in spin-coated $CH_3NH_3PbI_3$ perovskite thin film by transient fluorescent spectroscopy. A thickness-dependent fluorescent lifetime was found. By coating the film with an electron or hole transfer layer, [6,6]-phenyl-$C_{61}$-butyric acid methyl ester (PCBM) or 2,2',7,7'-tetrakis(N,N-di-p-methoxyphenylamine)-9,9'-spirobifluorene (Spiro-OMeTAD) respectively, we observed the charge transfer directly through the fluorescence quenching. One-dimensional diffusion model was applied to obtain long charge diffusion distances in thick films, which is ~1.7 μm for electrons and up to ~6.3 μm for holes. Short diffusion distance of few hundreds of nanosecond was also observed in thin films. This thickness dependent charge diffusion explained the formerly reported short charge diffusion distance (~100 nm) in films and resolved its confliction to thick working layer (300-500 nm) in real devices. This study presents direct support to the high performance perovskite solar cells and will benefit the devices' design.

Substantial attention has been drawn to the inorganic-organic perovskite-based solar cells, which currently achieve a certified high light conversion efficiency of 20.1%[1]. The combination of several excellent optoelectronic properties, such as very low exciton binding energy[2,3], highly mobile charge carriers[4-6], and efficient charge transportation to selective contact layers[3,5,7,8], makes perovskite "a game changer"[9] for photovoltaic devices and "a new avenue of research"[10]. As a fundamental issue, the carrier diffusion in perovskite is a major factor affecting the design and performance of the devices. However, this topic is still under debate at moment. It was shown that the charge diffusion distance in tri-iodine perovskite, $CH_3NH_3PbI_3$, is ~100 nm, studied by transient fluorescent spectroscopy[11,12]. On the other hand, many high efficient perovskite solar cells based on $CH_3NH_3PbI_3$ were made with perovskite layers thicker than this distance[13-15]. It is also investigated by impedance spectroscopy, photoinduced time-resolved microwave conductance (TRMC) and electron beam-induced current (EBIC) method, which hint a much longer charge transfer distance within perovskite layer[16-18]. A study on single crystal even give an extremely long diffusion length above 175 μm[19]. In addition, the diffusing balance between electrons and holes is not clear either. It was regarded that this balance is well maintained, while some reports say that the diffusion of holes are more/less efficient than electrons[17,20].

Beside the diffusion issue, some experimental observations are also in conflict. E.g. the fluorescent lifetime of the $CH_3NH_3PbI_3$ are dramatically varied in reports. In Xing's report, the lifetime is 4.5 ns[11], while in Stranks' report, it is 9.6 ns[12]. Some other experiments show that the lifetime for $CH_3NH_3PbI_3$ should be much longer than that. In reports by Yamada, the lifetime under low excitation light intensity can be 140 ns[21]. In single crystal, it is even longer than 100 μs under low excitation intensity[19]. This is an important parameter when calculating the charge diffusion distance by one-dimensional diffusion model[11,12]. It seems that all these conflicts need a better explanation.

To clarify these conflicts, we performed a study of directly observing the charge transfer in perovskite with various thicknesses and with an electron/hole transfer layer, by means of time-resolved transient fluorescence. It shows that the charge diffusion in $CH_3NH_3PbI_3$ is of distance at micrometer scale, which obviously longer than film thickness. The study also explains why former studies provide short diffusion lengths. The results show that hole transferring is faster than electron within perovskite thin film.

# RESULTS

**Absorption coefficient of $CH_3NH_3PbI_3$.** All perovskite films discussed here were prepared by a two-step dipping procedure similar to a report[22] and our study recently[23] on flat glass substrates. Figure 1 shows the absorption coefficient of $CH_3NH_3PbI_3$ derived from the absorption spectrum (see details in supplementary information, SI). This spectrum, which is in line with former reports, covers the entire UV and visible range up to 760 nm[11,24]. At 517 nm (the wavelength of pump light), a coefficient of $1.2 \times 10^5$ cm$^{-1}$ is slightly higher than the reference[11], corresponding to a penetration depth of 84 nm.

**Thickness dependence of lifetime.** The thickness of four prepared perovskite films are determined by a profilometer and listed in Table 1. An insulating polymer poly(methylmethacrylate) (PMMA) layer was coated atop the neat perovskite films for all photoluminescence (PL) decay measurement to passivate their moisture sensitivity[25]. By excitation at 517 nm, their transient fluorescent decay for the peak emission wavelength are shown in Fig. 2. The lifetimes for each thickness are also listed in Table 1. For a brief comparing, the curves are fitted by stretch exponential decay function[26]. The lifetimes show thickness dependency. For the films of 63 nm and 156 nm, their decays are 2.8 and 12.6 ns, which is similar to the reports[11,12]. For the two thick films of 254 and 310 nm, they have quite identical fluorescence decay as 90 ns and 91 ns. This means that the fluorescent decay is thickness-dependent in thin films, which disappears in thick ones.

**$CH_3NH_3PbI_3$ characterization.** The thickness-dependent fluorescence lifetime is analogous to previous reports on perovskite crystal grain size[26]. To investigate the influence of crystallite nature on PL properties, a series of scanning electron microscopy (SEM) and X-ray diffraction (XRD) measurement were performed on perovskite films of various thickness. Top-view SEM images of four samples mentioned above are shown in Fig. 3a-d. The thinnest film (made by 0.3 M $PbI_2$) in Fig. 3a is a thin layer of individual nanocrystallites with plenty of voids or pinholes. The average grain size is about few tens to ~100 nm. When films became thicker, as shown in Fig. 3b-d, the undesired voids evidently decreased, generating much more compact morphology. Meanwhile, larger crystallites were obtained in thick film, e.g. in Fig. 3d, the crystal size is of ~250 nm. This evolution of grain growth is in agreement with previous reported thermally annealed perovskite films[13,15]. The corresponding XRD patterns (Fig. 1 in SI) clearly show the perovskite structure (14.66°, 27.09°, 31.82°) with the presence of residual unreached $PbI_2$, in keeping with a previous study about $PbI_2$ deposited on flat glasses[22]. Moreover, the relative amount of $PbI_2$ decreases when thickness increases. Since the fabrication procedures for each sample are the same, this tendency should be attributed to the thickness for different samples.

**PL decays with and without quenchers.** To examine the charge transfer properties of the $CH_3NH_3PbI_3$ film, transient fluorescence experiments were performed by measuring the PL decay in perovskite film with or without a selected electron or hole acceptor. We made two samples with low and high $PbI_2$ concentrations, as listed in Table 2. Figure 4 shows the corresponding PL results of a thick perovskite film (390 nm). As shown in Fig. 4a and Table 2, the perovskite/PMMA film has a long lifetime of 170 ns, which we will explain in discussion. When the film was coated with a charge transfer layer, PCBM, e.g., fast fluorescent quenching happens (Fig. 4b). The decay is as fast as 1.24 ns, which means highly efficient electron transfer to the interface. It is even faster, when Spiro-OMeTAD is coated above perovskite films, as shown in Fig. 4c. The decay is 0.17 ns, close to the instrument response of streak camera. For the thinner film of 95 nm, PL decays show the same trend (Fig. 2 in SI). The neat perovskite film has a lifetime of 12.4 ns, which decreases to 0.40 ns and 0.16 ns for PCBM and Spiro-OMeTAD coated samples, respectively.

**DISCUSSION**

The thickness-dependent lifetimes indicate that the fluorescent quenching is neither local, nor to the surfaces. Then this quenching is more like a boundary related effect. We observed that the size of the grain become larger when the films are thicker. In addition, the films become compact with less defects. Then the abundant surface area, void, and defects in thin film should be responsible for the quenching. In thick films, the boundary effect become insignificant. In our study, it is above 250 nm, as shown in Fig. 2. This grain size independency was also shown in D'Innocenzo's report, in which the lifetime of ~100 ns was found for grains between 0.2-2 micron[26]. Smaller grains present dramatically reduced lifetime was also presented. This proves that when the grain size become too small, it will produce reduced lifetime. However, in an optimized real device of 300-500 nm, the small size grain is avoided and the lifetime is not sensitive to detailed morphology. On the contrary, in semi-transparent device and thin films for photophysical study, small size grain exists with reduced lifetime. Therefore, towards its real applications, we prefer to take film thickness as a basic parameter to describe the lifetime dependency, instead of grain size and defects, which had already been optimized in cells with high efficiency by many groups.

The SEM and XRD studies reveal the mechanism of this dependency. The SEM images manifest the evolution of grain growth, from the level of below 100 nm to ~250 nm, and the diminishing of voids or pinholes. We can rationally assume that long lifetime exists when large size of crystallite are the main species in film, and the boundaries between crystallites have a significant impact on the fluorescence characteristic[27,28], esp. when the crystal size is small and not compact. Some groups showed that a proper amount of $PbI_2$ species can fill perovskite grain boundaries, eliminate defect states, and thus slow down the carrier relaxation, whereas a large amount of excessive $PbI_2$ is detrimental to charge transport[29,30]. This is in according with our XRD results for thick and thinner perovskite films, respectively. We believe this is the quenching mechanism of thickness dependent fluorescent lifetime. In short, when the perovskite layer become thicker, it has larger crystallite size with reduce overall grain boundary area, and much less defects due to reduced $PbI_2$ at boundary. Both the factors finally make the fluorescence emitted by thick film independent to thickness.

It has been well established that excitons in perovskite are nearly fully ionized because of low binding energy[2,31,32]. So the charge diffusion directly relate to their lifetime. Therefore, the film thickness dependent fluorescent lifetime becomes an important issue here. When the film is thin, the lifetime is short due to the boundary defects. This means that to find out the unaffected charge diffusion distance, the real lifetime needs to be established in advance. As mentioned earlier, in the thicker film of ~280 nm, a long lifetime of 140 ns can be found at lowest pump intensity[21]. In addition, when the perovskite are in large crystal of ~1 μm, its lifetime is also at ~100 ns timescale[26]. These results are very similar to our observation in thick films. It reasonably suggests that the native lifetime for $CH_3NH_3PbI_3$ without considerable boundary defect is at ~100 ns timescale, though small differences exist among research groups. To our knowledge, for reports whose lifetime is ~100 ns, they are shown as thicker films or larger grain sizes.

We performed a step forward experiment to verify the grain size dependent fluorescent decay. For thick film of 345-390 nm, grain size of 150-350 nm were prepared, shown in SI Fig. 4. We found that grain size has little effects on lifetime. All of them present lifetime of ~200 ns, as shown in SI Fig. 5 and SI Table 1. Therefore, we can conclude that the grain size should have no significant effect on charge diffusion. The grain size has large tunable range when it is compact and with less defects. This also explains that high performance perovskite solar cells can be repeated among labs, in spite of morphological variation.

The one-dimensional diffusion model are describe in the SI. The fittings produce diffusion constant, $D$. The charge diffusion distance $L_D$, is calculated by the equation $L_D = \sqrt{D\tau_{PL}}$, where $\tau_{PL}$ is PL lifetime of 390 nm in this study by simple rate equation[21,31,32]. We take the duration when fluorescence decays to 1/e of initial intensity as the diffusion time for electrons and holes. As summarized in Table 2, we obtain the electron diffusion coefficient of 0.18 cm$^2$ s$^{-1}$ and corresponding diffusion length of ~1.7 μm. This confirms the observation by EBIC method[17]. As a comparing, when taking lifetime of the thinner film, 95 nm, we can calculate the corresponding diffusion constant of 0.06 cm$^2$ s$^{-1}$ and diffusion length of 273 nm. This result is close to other reports based on thin films[11,12,33,34]. For diffusion of hole, the $D$ is found 2.3 cm$^2$ s$^{-1}$ in perovskite/Spiro-OMeTAD film, which is one order larger than the electron diffusion coefficient. The corresponding charge diffusion distance is calculated as ~6.3 μm. For the thin film of 95 nm, this distance is 459 nm, as listed in Table 2. It should be remarked here that both the electron and hole's diffusion distances obtained are much longer than the thickness of the perovskite films ever made with top solar energy conversion efficiency.

There are several points should be addressed here. The first is that though the charge diffusion distance are thickness dependent, it is much longer than the film thickness, even for the thin film less than 100 nm. Therefore, both in thin semi-transparent devices and black thick devices, the high cell performance can both be achieved. The long diffusion distance provide a large tunable range for preparing perovskite working layers, both for thickness and morphology. The second is that the charge transfer balance between the electrons and holes is not exactly shown in our study.

However, it is less important for currently developed devices, which are usually 300-500 nm. The last one is that the lifetime of $CH_3NH_3PbI_3$ may not be an exact number but a range around 100 ns or larger, since the crystal grain size, defects, preparation procedure, post treatment, pump energy, and fitting methods *et. al.* can not exactly be the same. However, the lifetime variation will not make the diffusion distance lower than micrometer level.

The plain film provide a simple model to study the charge diffusion inside perovskite working layers. Another widely applied cell structure is with mesoporous scaffold such as $TiO_2$ and $Al_2O_3$. Due to the restricted growth for crystals and large interface area, the charge diffusion inside the structure is much more complicated. E.g. the fluorescent lifetime varies to the size of mesoscopic pores, the type of scaffold, and capping layer.[35, 36] High speed charge transfer to $TiO_2$ of 200 fs was also found through ultrafast spectroscopy.[37] At molecular level, the interface shows oriented permanent dipoles indicating the existence of ordered perovskite layer.[38] In addition, the perovskite inside scaffold shows co-existence of medium range crystalline and local structural coherence.[39] These studies reveal the significant differences of these mesoporous cells compared to the planner structured devices. In spite of these differences, it has been reported that the charge diffusion in mesoporous devices has lower coefficient than the layered cells.[40] It is also reported that the existence of capping layer may recover the longer charge diffusion distance.[36] Base on these study, we can summarize that the charge diffusion distance inside the mesoporous layer is smaller, but may benefit from the large interface area for high efficient charge extraction.[5]

In conclusion, we found that the fluorescent lifetime of spin-coated $CH_3NH_3PbI_3$ perovskite thin film depends on the film thickness. The lifetime increases towards the increment of film thickness, till ~250 nm. The lifetime finally increases to >100 ns. The fluorescent quenching in thin film is due to the defects at grain boundary. Therefore we take a thick film of 390 nm to study the charge diffusion in $CH_3NH_3PbI_3$. After coating charge-transfer layer, PCBM or Spiro-OMeTAD, and applying one-dimensional diffusion model, we can obtain the charge diffusion distance of 1.7 μm for electrons and 6.3 μm for holes. For thin film of 95 nm, a result of short diffusion distance similar to other reports are found. This study resolves the current conflict between the measured short charge diffusion distance and thick working layer in high efficient devices. The result shows that in case of thin and thick films, $CH_3NH_3PbI_3$ both can provide long charge diffusion distance for best cell performance.

**METHODS**

**Sample preparation**. All samples were fabricated on glass slide substrates. First, glass substrates were cleaned sequentially by ultrasonic bath in detergent water, deionized water, acetone and ethanol for 15 min, respectively, and then exposed to oxygen plasma for 15min to achieve optically smooth films. The $CH_3NH_3PbI_3$ perovskite were fabricated with a two-step sequential deposition method under nitrogen atmosphere. The pre-cleaned glass substrates were spin-coated a $PbI_2$ solution (6000 rpm, 0.3 M, 0.5 M, 0.8 M, and 1.1 M) of *N,N*-dimethylformamide (DMF) at ambient temperature to obtain layers of different thicknesses. After drying at 60 ℃ in ambient environment for 6 h, the films were dipped into $CH_3NH_3I$ solution in 2-propanol (15 mg/mL) at 65 ℃ for 90 s and then rinsed with 2-propanol. For the samples fabricated by using various concentration of $CH_3NH_3I$ solution (10 mg/mL, 15 mg/mL and 20 mg/mL), $PbI_2$ was kept at 1.0 M. After $CH_3NH_3PbI_3$ annealing at 100 ℃ for 40 min, Spiro-OMeTAD (10 mg/mL), PCBM (10 mg/mL) or PMMA (20 mg/mL) was spin-coated at 2000 rpm for 60 s atop the $CH_3NH_3PbI_3$ perovskite films.

**Characterization details**. XRD patterns were obtained using a Philips X'PERT-MRD x-ray diffractometer system with a Cu Kα radiation source (λ=0.1541 nm) at 45 kV and 40 mA. SEM images were collected using a Hitachi S-4800 microscope, with a working bias of 10 KV. Sample thicknesses were measured using a Veeco Dektak 150 profilometer. Ultraviolet-visible (UV-vis) absorption measurements were recorded with an Agilent 8453 UV–vis Spectroscopy System at room temperature.

**Time-resolved photoluminescence.** The time-resolved fluorescence spectra were recorded with a high resolution streak camera system (Hamamatsu C10910). We used an amplified mode-lock Ti: Sapphire femtosecond laser system (Legend, Coherent) and a two-stage optical parametric amplifier (OperA Solo, Coherent) to generate the pump beam with a repetition rate of 1 KHz. All the samples were excited by 517 nm at room temperature. The excitation fluence on the sample surface was in the range from 9 nJ/cm$^2$ to 1.2 μJ/cm$^2$ per pulse.

**Acknowledgements**

This work supported by the National Basic Research Program of China 2013CB921904, 2011CB933303; National Natural Science Foundation of China under grant Nos. 61177020, 11134001, 11574009, 61575005, 21321001, 21371012.

**Author contributions**

S.W. and Z.B. proposed the idea and designed the experiments. Y.L. performed the PL, UV-vis, XRD measurement, and data analysis. W.Y. fabricated all samples. Y.L. did SEM. W.W. contributes in transient PL study. L.X. and Q.G. gave helpful discussion and contributed in manuscript writing.

**Competing financial interests:** The authors declare no competing financial interests.

**FIGURES**

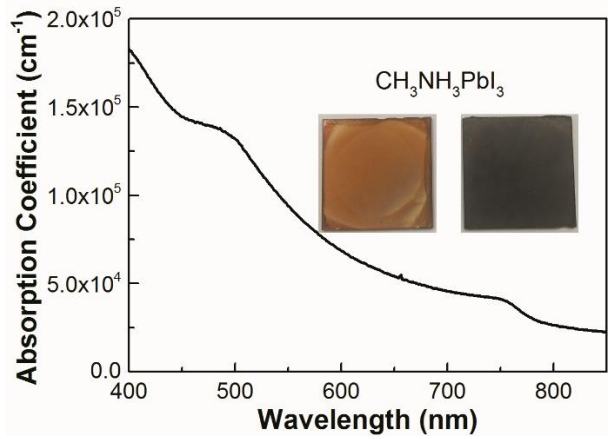

**Figure 1.** Plot of absorption coefficient. Absorption coefficient towards wavelength for $CH_3NH_3PbI_3$ thin film prepared via a two-step deposition method. Insets are the top view photos of perovskite samples of a thin (yellow brown, left) and a thick (dark brown, right) one on glass substrates.

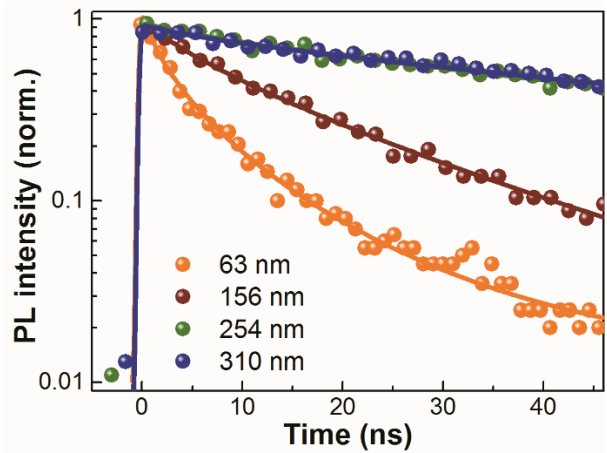

**Figure 2.** Thickness-dependent time-resolved PL data. PL decay curves of $CH_3NH_3PbI_3$ of different thicknesses depending on varied $PbI_2$ concentration (63 nm, 0.3 M; 156 nm, 0.5 M; 254 nm, 0.8 M; and 310 nm, 1.1 M, respectively) upon excitation at 517 nm, 90 nJ/cm$^2$. The solid lines are the stretched exponential fits to the corresponding results.

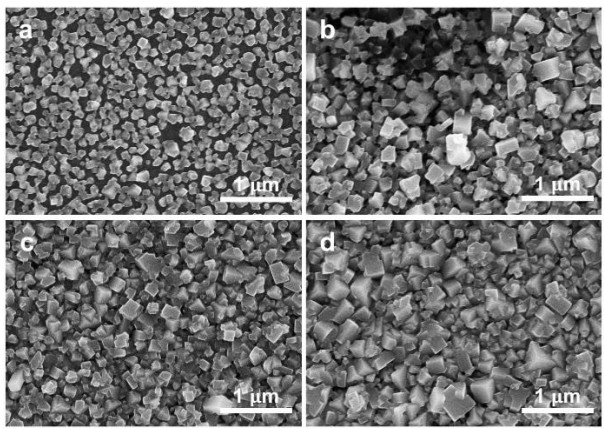

**Figure 3.** SEM images. Top-view SEM images of CH$_3$NH$_3$PbI$_3$ deposited via changing PbI$_2$ precursor concentration. (**a**) 0.3 M; (**b**) 0.5 M; (**c**) 0.8 M; and (**d**) 1.1 M.

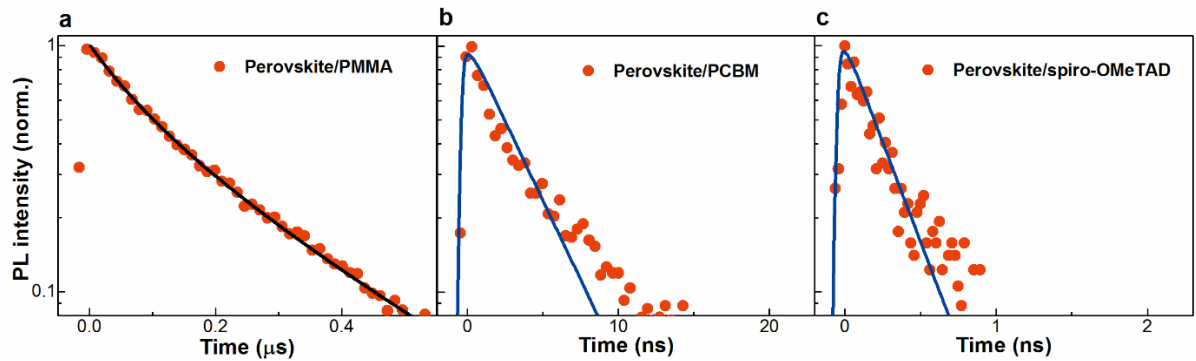

**Figure 4.** Time-resolved PL decays of perovskite $CH_3NH_3PbI_3$ coated with different layers (red circles). (**a**) spin-coated $CH_3NH_3PbI_3$ perovskite film (390 nm); (**b**) film covered by PCBM; and (**c**) covered by Spiro-OMeTAD, excited at 517 nm, 90 nJ/cm$^2$. The black and blue solid lines are the fits to the PL results by rate equation and one-dimensional diffusion model, respectively.

# TABLES

**Table 1.** Summary of the stretched exponential fitting for the PL decays in Figure 2.

| Concentration of PbI$_2$ (M) | Thickness (nm) | $\tau_s$ (ns) | $\beta$ |
|:---:|:---:|:---:|:---:|
| 0.3 | 63 | 2.8 | 0.57 |
| 0.5 | 156 | 12.6 | 0.74 |
| 0.8 | 254 | 90 | 0.64 |
| 1.1 | 310 | 91 | 0.64 |

**Table 2. Summary of parameters in Figure 4**. the fluorescent decay parameters obtained by rate equation (the monomolecular trapping rate ($A$), bimolecular radiative recombination coefficient ($B$), and effective PL lifetime ($\tau_{PL}$)), the PL decay of perovskite coated with and an electron transport layer (ETL, PCBM) and hole transport layer (HTL, Spiro-OMeTAD), the calculated diffusion coefficients ($D$), and diffusion lengths ($L_D$) of a thin and a thick $CH_3NH_3PbI_3$ perovskite films.

| Thickness (nm) | $A$ (s$^{-1}$) | $B$ (cm$^3$/s) | $\tau_{PL}$ (ns) | $\tau_{w/ETL}$ (ns) | $\tau_{w/HTL}$ (ns) | Species | $D$ (cm$^2$/s) | $L_D$ (µm) |
|---|---|---|---|---|---|---|---|---|
| ~95 | 2.8×10$^7$ | 2.2×10$^{-9}$ | 12.4 | 0.40 | 0.16 | Electrons | 0.06 | 0.27 |
| | | | | | | Holes | 0.17 | 0.46 |
| ~390 | 3.0×10$^6$ | 5.1×10$^{-10}$ | 170 | 1.24 | 0.17 | Electrons | 0.18 | 1.7 |
| | | | | | | Holes | 2.3 | 6.3 |

## Supplementary Methods

**Calculation of absorption coefficient.** The absorption (*A*) was calculated using:

$$A = -log(\frac{I_s}{I_0}) \tag{S1}$$

where $I_0$ is the reference intensity measured with the blank and $I_s$ is the intensity measured with the sample.

The absorption coefficient α was obtained from:

$$T = \frac{I_s}{I_0} = exp(-\alpha L) \tag{S2}$$

where *T* represents transmittance and *L* is the sample thickness.

From Equation S1 and S2, we can get the absorption coefficient by the value of *A*:

$$\alpha = A \cdot ln10/L \tag{S3}$$

The absorption coefficient for CH$_3$NH$_3$PbI$_3$ is as a function of wavelength. The penetration depth $\tau(\lambda)$ is then calculated as $\tau(\lambda) = 1/\alpha(\lambda)$.

**Calculation of charge diffusion coefficient and diffusion length**.

The charge diffusion length was calculated based on one-dimensional diffusion model which was described in Xing's paper (Xing, G. *et al. Science* **342,** 344-347 (2013)) as below:

$$N(t) = \frac{2n_0 L}{\pi} exp(-k(t)t) \sum_{m=0}^{\infty} (exp\left(-\frac{\pi^2 D}{L^2}\left(m+\frac{1}{2}\right)^2 t\right) \frac{exp(-\alpha L)\pi\left(m+\frac{1}{2}\right)+(-1)^m \alpha L}{\left((\alpha L)^2+\pi^2\left(m+\frac{1}{2}\right)^2\right)\left(m+\frac{1}{2}\right)}) \tag{S4}$$

where $N(t)$ is overall photocarrier density within perovskite film, $D$ is the diffusion coefficient, $n_0$ is initial photoinduced carrier distribution, $k(t)$ is the observed PL decay rate without any quencher layers, α is the absorption coefficient, and $L$ is the thickness of perovskite layer.

By fitting time-resolved PL results we can obtain the expression of $k(t)$. PL decay curves in our experiments are fitted by stretched exponential model or simple rate equation.

For stretched exponential fitting:

$$I(t) = I_0 e^{-(t/\tau_s)^\beta} \tag{S5}$$

From Equation (S5) we can obtain the expression of $k(t) = \beta \tau_s^{-\beta} t^{\beta-1}$.

For pristine films with mono- and bi-molecular behavior, we use simple rate equation:

$$-\frac{dn}{dt} = An + Bn^2 \tag{S6}$$

$$I_{PL} \propto Bn^2 + BNn \tag{S7}$$

where $n$ is the photogenerated carrier density, $A$ is the monomolecular trapping rate, $B$ is the bimolecular radiative recombination coefficient and $N$ is the emissive trap density. The effective PL lifetime then can be written as

$$\tau_{PL} = (A + Bn_0)^{-1} \tag{S8}$$

From Equation (S8) we obtain $k(t) = (A + Bn_0)$. Sequentially $N(t)$ is used to fit the PL decay of perovskite films coated with quenchers to derive the value of $D$. It is notable that the observed

PL intensity $I(t)$ is the convolution of the intrinsic PL intensity $f(t)$ and the instrument response function (IRF) $g(t)$:

$$I(t) = \int g(t)f(t-t')dt' \tag{S9}$$

So by fitting the observed PL decay with the convolution of IRF gives the estimated $D$ value. Finally the charge carrier diffusion length $L_D$ is calculated by $L_D = \sqrt{D\tau_{PL}}$, where $\tau_{PL}$ is the fitted PL lifetime in the case of no quenching layers.

**Supplementary Figure 1.**

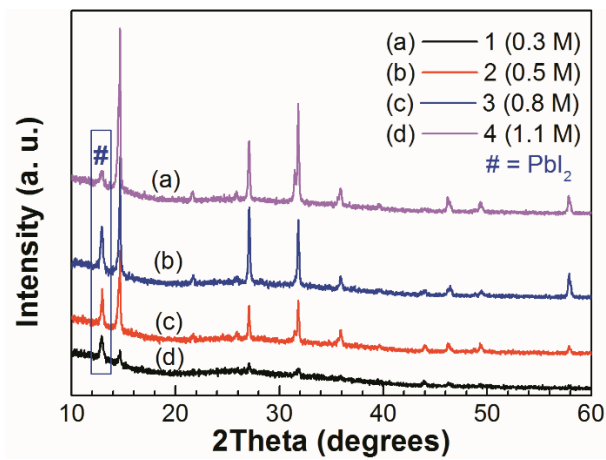

**Supplementary Figure 1.** XRD patterns of CH$_3$NH$_3$PbI$_3$ films fabricated with varied PbI2 concentration: (**a**) 0.3 M, (**b**) 0.5 M, (**c**) 0.8 M, and (**d**) 1.1 M, with corresponding thickness of 63, 156, 254, and 310 nm.

**Supplementary Figure 2.**

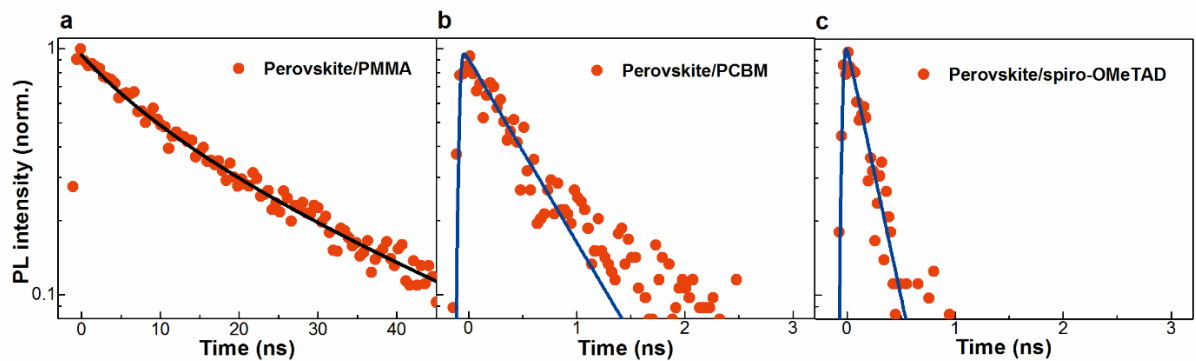

**Supplementary Figure 2.** PL decays (red circles) of $CH_3NH_3PbI_3$ perovskite (ca. 95 nm) coated with (**a**) PMMA, (**b**) PCBM and (**c**) Spiro-OMeTAD layer, taken at the peak emission wavelength, excited at 517 nm, 90 nJ/cm$^2$. The black (a) and blue solid lines (b and c) are the fits to the PL results by using rate equation and one-dimensional diffusion model, respectively.

**Supplementary Figure 3.**

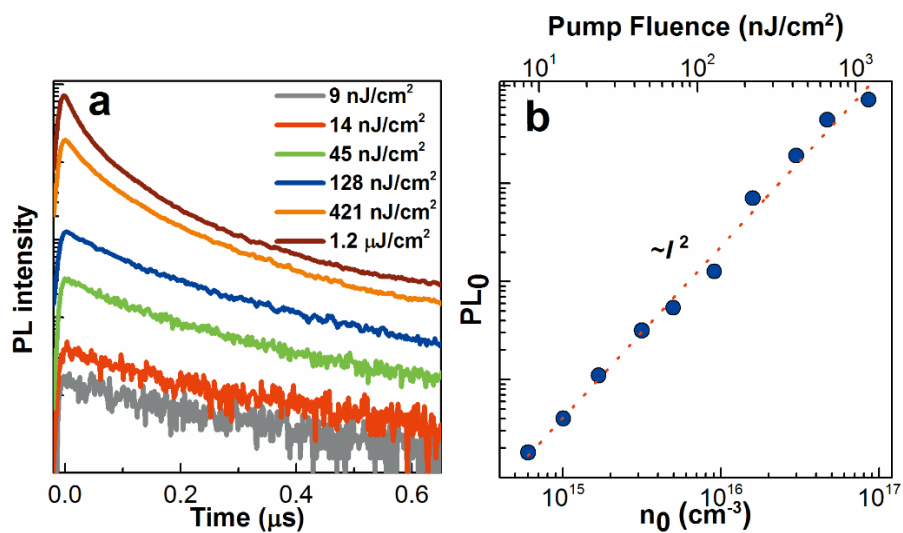

**Supplementary Figure 3.** Time-resolved photoluminescence spectroscopy and their pump energy dependency. (a) Pump fluence dependent PL decays of a thick $CH_3NH_3PbI_3$ perovskite (ca. 370 nm) thin film, excited at 517 nm. (b) Photoluminescence emission intensity at t=0 ($PL_0$) as a function of photocarrier density (lower axis) and laser pump fluence (upper axis).

**Supplementary Figure 4.**

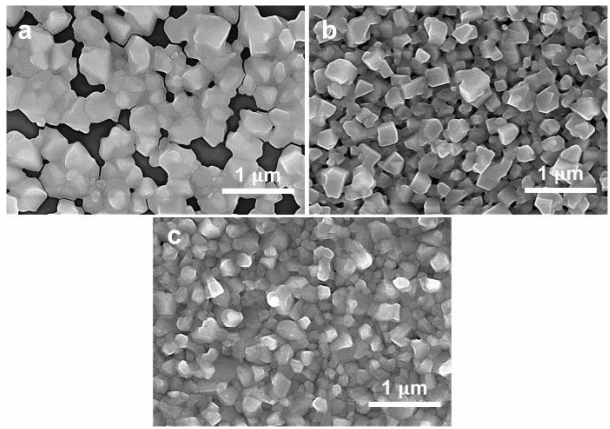

**Supplementary Figure 4.** SEM images of a series of thick perovskite film containing crystal grains of different sizes. Top-view SEM images of $CH_3NH_3PbI_3$ deposited via changing $CH_3NH_3I$ concentration: (**a**) 10 mg/mL; (**b**) 15 mg/mL; (**c**) 20 mg/mL.

**Supplementary Figure 5.**

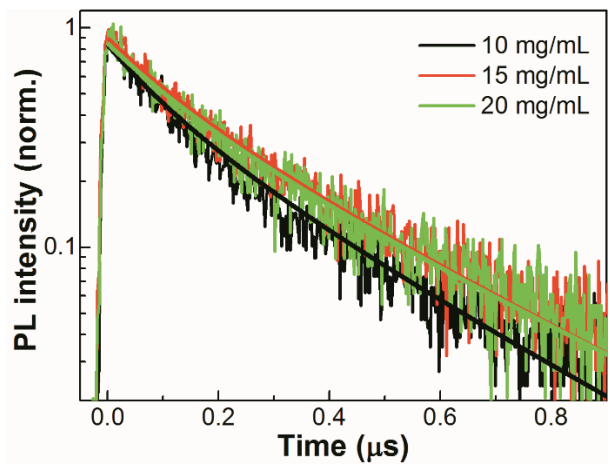

**Supplementary Figure 5.** PL decay curves of $CH_3NH_3PbI_3$ thin films made by various $CH_3NH_3I$ concentration upon excitation at 517 nm, 24 nJ/cm$^2$. The solid lines are the rate equation fits to the corresponding results.

**Supplementary Figure 6.**

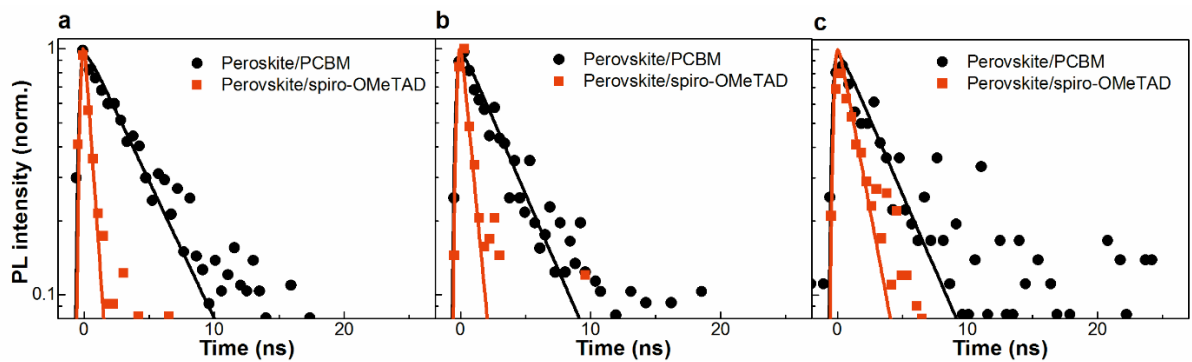

**Supplementary Figure 6.** PL decay curves of $CH_3NH_3PbI_3$ thin films coated with an electron (PCBM; black circles) or hole (Spiro-OMeTAD; red squares) quenching layer upon excitation at 517 nm, 24 $nJ/cm^2$. $CH_3NH_3PbI_3$ thin films are formed using varied $CH_3NH_3I$ concentrations shown in SI Figure 4 &5: (**a**) 10 mg/mL; (**b**) 15 mg/mL; (**c**) 20 mg/mL. The solid lines are the one-dimensional model fits to the corresponding results.

**Supplementary Table 1.** Summary of the thickness ($L$), the fitted monomolecular trapping rate ($A$), bimolecular radiative recombination coefficient ($B$), the trap density ($N$) and the effective PL lifetime ($\tau_{PL}$) in Supplementary Figure 5. The lifetime of perovskite coated with electron transport layer (ETL) or hole transport layer (HTL), the calculated diffusion coefficients ($D$), and diffusion lengths ($L_D$) in Supplementary Figure 6 are also presented.

| $CH_3NH_3I$ (mg/mL) | $L$ (nm) | $A$ (s$^{-1}$) | $B$ (cm$^3$/s) | $N$ (cm$^{-3}$) | $\tau_{PL}$ (ns) | $\tau_{w/ETL}$ (ns) | $\tau_{w/HTL}$ (ns) | Species | $D$ (cm$^2$/s) | $L_D$ (μm) |
|---|---|---|---|---|---|---|---|---|---|---|
| 10 | ~345 | 3.2×10$^6$ | 1.2×10$^{-9}$ | 4.0×10$^{15}$ | 189.0 | 3.1 | 0.57 | Electrons | 0.12 | 1.5 |
| | | | | | | | | Holes | 0.85 | 4.0 |
| 15 | ~370 | 2.9×10$^6$ | 0.9×10$^{-9}$ | 3.7×10$^{15}$ | 230.0 | 2.8 | 0.63 | Electrons | 0.15 | 1.7 |
| | | | | | | | | Holes | 0.70 | 4.0 |
| 20 | ~390 | 2.8×10$^6$ | 1.0×10$^{-9}$ | 4.4×10$^{15}$ | 222.8 | 2.5 | 1.2 | Electrons | 0.17 | 1.9 |
| | | | | | | | | Holes | 0.40 | 3.0 |

**Supplementary Table 2.** Pump fluence dependent PL lifetime. The effective PL lifetime for $CH_3NH_3PbI_3$ thin films fabricated by using 1.0 M $PbI_2$ and various concentration of $CH_3NH_3I$ solution (10 mg/mL, 15 mg/mL, and 20 mg/mL).

| Pump Fluence (nJ/cm$^2$) | $\tau_{PL}$ (ns) | | |
| --- | --- | --- | --- |
| | 10 mg/mL | 15 mg/mL | 20 mg/mL |
| 9 | 254.8 | 295.5 | 293.4 |
| 14 | 225.4 | 266.9 | 262.3 |
| 24 | 189.0 | 230.0 | 222.8 |
| 45 | 139.7 | 176.4 | 167.6 |
| 70 | 105.7 | 137.1 | 128.4 |